\documentclass[]{piparticle-final}
\usepackage{graphicx}
\usepackage{amsmath}
\usepackage{cite}

\begin{document}

\volume{2}               
\articlenumber{020006}   
\journalyear{2010}       
\editor{I. Ippolito}   
\reviewers{A. Coniglio, Universit\'{a} di Napoli ``Federico II'', Napoli, Italy.}  
\received{10 August 2010}     
\accepted{28 October 2010}   
\runningauthor{T. Divoux}  
\doi{020006}         

\title{Invited review: Effect of temperature on a granular pile}

\author{Thibaut Divoux,\cite{inst1}\thanks{E-mail: Thibaut.Divoux@ens-lyon.fr}}

\pipabstract{
As a fragile construction, a granular pile is very sensitive to minute external perturbations. In particular, it is now well established that a granular assembly is sensitive to variations of temperature. Such variations can produce localized rearrangements as well as global static avalanches inside a pile. In this review, we sum up the various observations that have been made concerning the effect of temperature on a granular assembly. In particular, we dwell on the way controlled variations of temperature have been employed to generate the compaction of a granular pile. After laying emphasis on the key features of this compaction process, we compare it to the classic vibration-induced compaction. Finally, we also review other \textit{granular systems} in a large sense, from microscopic (jammed multilamellar vesicles) to macroscopic scales (stone heave phenomenon linked to freezing and thawing of soils) for which periodic variations of temperature could play a key role in the dynamics at stake.
}

\maketitle

\blfootnote{
\begin{theaffiliation}{99}
   \institution{inst1} Universit\'e de Lyon, Laboratoire de Physique, \'Ecole Normale Sup\'erieure de Lyon, CNRS UMR 5672, 46 All\'ee d'Italie, 69364 Lyon cedex 07, France.
\end{theaffiliation}
}

\section{Introduction: a granular pile as a fragile construction}
\label{intro}

\begin{figure}[t]
\begin{center}
\includegraphics[width=0.45\textwidth]{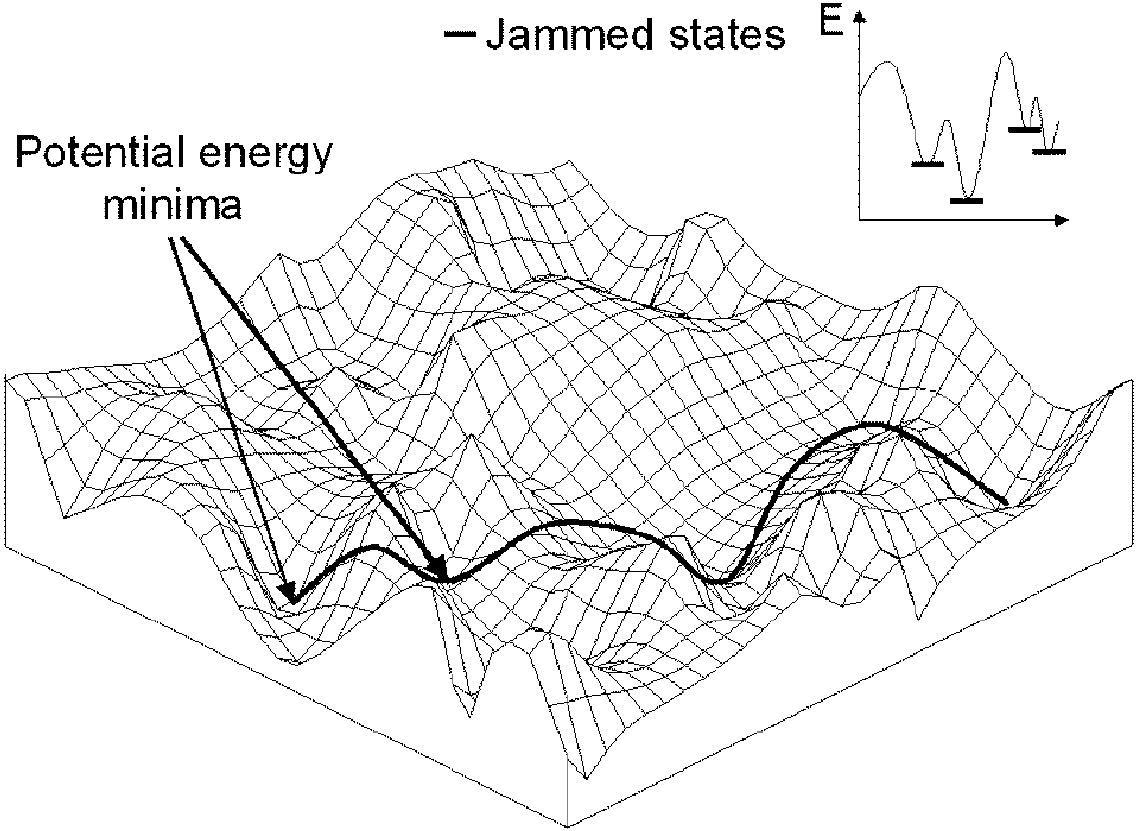}
\end{center}
\caption{Sketch of the energy landscape associated to the dynamics of a glass, below the glass transition. In the case of a granular pile, the thermal agitation is negligible and one has to inject energy to overcome the energy barrier separating the jammed states. Reproduced with permission from \cite{Makse2004}. Copyright Wiley-VCH Verlag GmbH \& Co. KGaA.} \label{fig.Energylandscape}
\end{figure}

\begin{figure}[b]
\begin{center}
\includegraphics[width=0.45\textwidth]{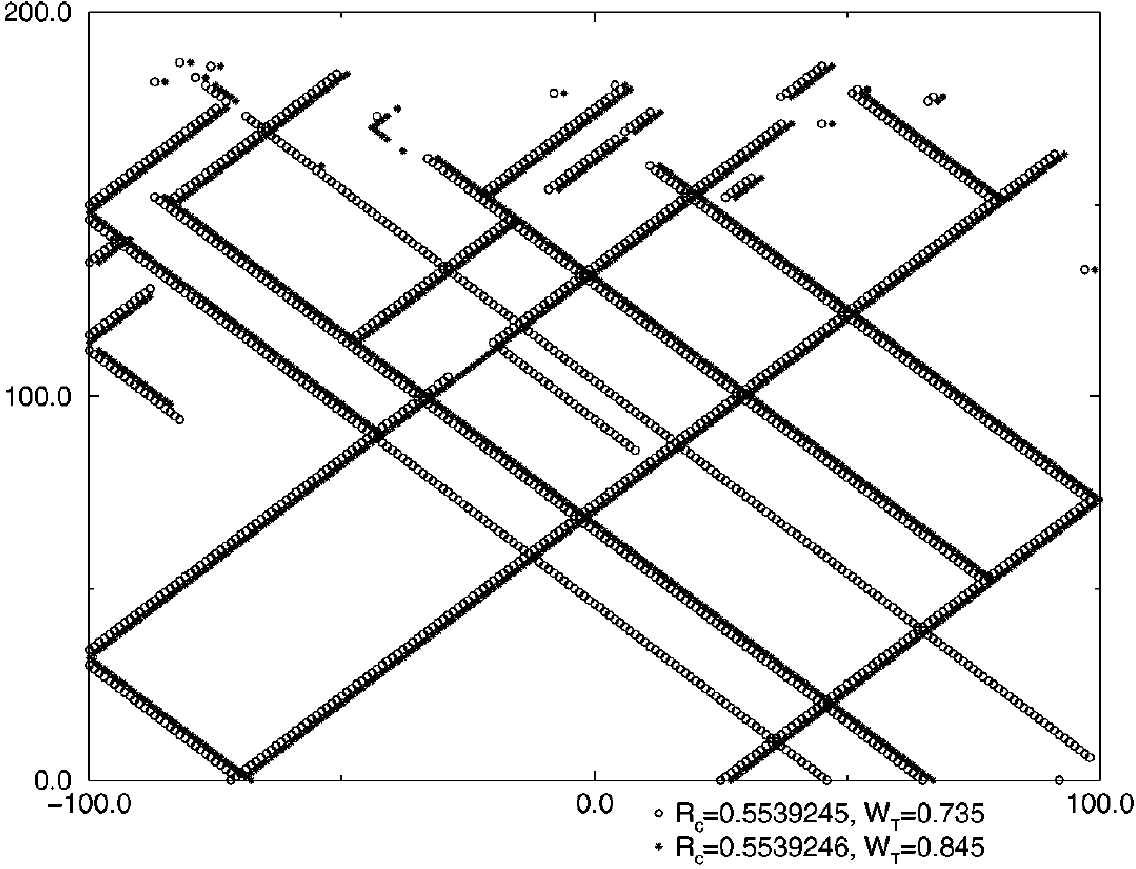}
\end{center}
\caption{Force chains in a two-dimensional granular pile (200$\times$200) following the Scalar Arching Model. The white dots label the grains involved in a force chain in the initial force network, whereas the black dots label the grains involved in the force chains network after a relative change of $10^{-7}$ of the friction coefficient $R_c$. One observes that the reorganization takes place in the whole pile, despite the perturbation taking place at the scale of the grain surface roughness. Reprinted figure with permission from \cite{Claudin1997}. Copyright (1997) by the American Physical Society.} \label{fig.SAM}
\end{figure}

A granular pile can be described as a bunch of hard and frictional grains for which the thermal ambient agitation is negligible \cite{Weeks2007}. Indeed, the potential energy of a grain of density $\rho$ assessed over a displacement equivalent to its diameter $d$ satisfies $\rho g d^4/k_BT \simeq 10^{11} \gg 1$. Thus a granular pile is an athermal system, and one needs to inject energy inside the pile to trigger any reorganization of the packing. The accessible packings can then be seen as jammed states, i.e. minima of the potential energy, in some energy landscape (Fig.~\ref{fig.Energylandscape}) and the energy one injects makes it possible to overcome the energy barrier that separates two configurations from one another. Various methods have been used, over the passed 20 years, to provide energy to a granular pile, among which mechanical vibrations \cite{Knight1995,Kudrolli2004,Richard2005} and shear \cite{Howell1999,Toiya2004} are the most widespread.
Nevertheless, this description of a granular pile in terms of an athermal system hinders one of its major characteristics: namely its {\it fragility} \cite{Weeks2007,Cates1998}. The sensitivity of a granular pile to minute external perturbations has first been pointed out in the case of frictionless hard spheres \cite{Ouagenouni1995,Ouagenouni1997,Moukarzel1998a,Moukarzel1998b}. A heap of such spheres presenting a slight polydispersity has been shown to be isostatic and thus very sensitive to external perturbations. In the case of frictional spheres, jamming and isostaticity no longer go hand in hand and the way the packing has been build plays a crucial role on its stability \cite{VanHecke2010}. Nevertheless, the pile pictured as a contact network, can be decomposed in two subnetworks: a network gathering {\it strong contacts} ({\it weak contacts} resp.) involving grains which carry a force larger (lower resp.) than the average force in the packing \cite{Radjai1998}. The key contribution of both this ``strong contact" network and the surface roughness of the grains to the fragility of the pile is very well illustrated by the Scalar Arching Model (SAM) \cite{Claudin1997}. This model, inspired from the q-model developed by Liu, Coppersmith et al. \cite{Liu1995,Coppersmith1996} only takes into account the weight of the grains and the solid friction between the grains following Coulomb's law. The control parameter of the simulation is the friction coefficient between two grains, denoted $R_c$. In this model, looking at a static pile, Claudin and Bouchaud demonstrated that a relative variation of $R_c$ as small as $10^{-7}$ triggers large scale reorganizations inside the pile, named static avalanches (Fig.~\ref{fig.SAM}), emphasizing the fragility of a pile to minute perturbations, despite its athermal nature \cite{Claudin1997}. 
Of course, in a laboratory experiment, controlling and varying the friction coefficient between the grains during an experiment is impossible. Nonetheless, as suggested in \cite{Claudin1997}, variations of temperature might perturb the pile at the scale of the surface roughness of the grains, having an effect equivalent to a change of the friction coefficient. Indeed, one can assess that a granular heap of size $L$
(typically a few centimeters) submitted to variations of temperature of amplitude $\Delta T$ experiences a dilation $\delta L=\kappa_g L \Delta T$, where $\kappa_g$ stands for the thermal expansion coefficient of the grains. Dilations corresponding to the surface roughness scale lead to $\Delta T \simeq 0.1^{\circ}$C for standard glass beads ($\kappa_g=10^{-6}$~K$^{-1}$). Such an amplitude is easily accessible and for instance daily variations of temperature in the lab are already of roughly a few degrees. This is the topic of this brief review, where we focus on the effect of temperature on a granular assembly. The content of this review goes as follows: In section~II we sum up the first experimental observations of uncontrolled temperature variations that have pushed for further experiments under controlled variations (section~III). In section~IV, we dwell on the use of cycles of temperature to induce the compaction of a granular pile in a delicate way, in particular we discuss the role of both the amplitude and the frequency of the imposed cycles. Section~V deals with enlarged ``granular" systems and extend the scope of the results presented in previous parts. Finally, section~VI proposes some outlooks to this method of thermal cycling as a general method to probe granular systems.

\section{From undesired variations of temperature ...}
\label{accident}

The effect of temperature variations over a granular pile has first been reported as a hindrance to perform reproducible measurements. Those experiments were not dedicated to probe the consequences of temperature variations on a granular assembly, and the role of temperature is simply assessed. Nonetheless, several issues were raised in this seminal work, and are highlighted here.

\subsection{Sound in Sand}

The influence of temperature is first mentioned in the early 90's in a work dealing with sound propagation in sand \cite{Liu1992,Liu1994,Liu1994b}. An acoustic wave is generated inside a granular pile and recorded a few centimeters away. C. Liu and S. Nagel observed that ``\textit{a temperature change of only 0.04 K inside the pile, produced by the change of the ambient temperature, or by a local heater, could cause a factor of 3 reversible change in the measured vibration transmission}" \cite{Liu1992}. One can indeed assess the grain dilation $\delta d$ associated to such variations of temperature $\Delta T$ to be $\delta d = \kappa_g \Delta Td \simeq 2$~nm. In agreements with the reversibility of the effect, such an amplitude is on the one hand negligible compared to the typical surface roughness of the beads used in their experiments (probably 100~nm for glass beads \cite{Divoux2009b}) and, on the other hand less than the typical deformation of the beads inside the heap (roughly 10~nm assuming a Hertz-like contact). However, at the time of these experiments, it was still unclear whether the temperature or the gradient of temperature were leading to such observations. Experiments performed by Cl\'ement and co-workers later confirmed the key role of the gradient, and brought to the fore the role of the container dilation on the reorganization process \cite{Clement1997}.

\begin{figure}[th]
\begin{center}
\includegraphics[width=0.45\textwidth]{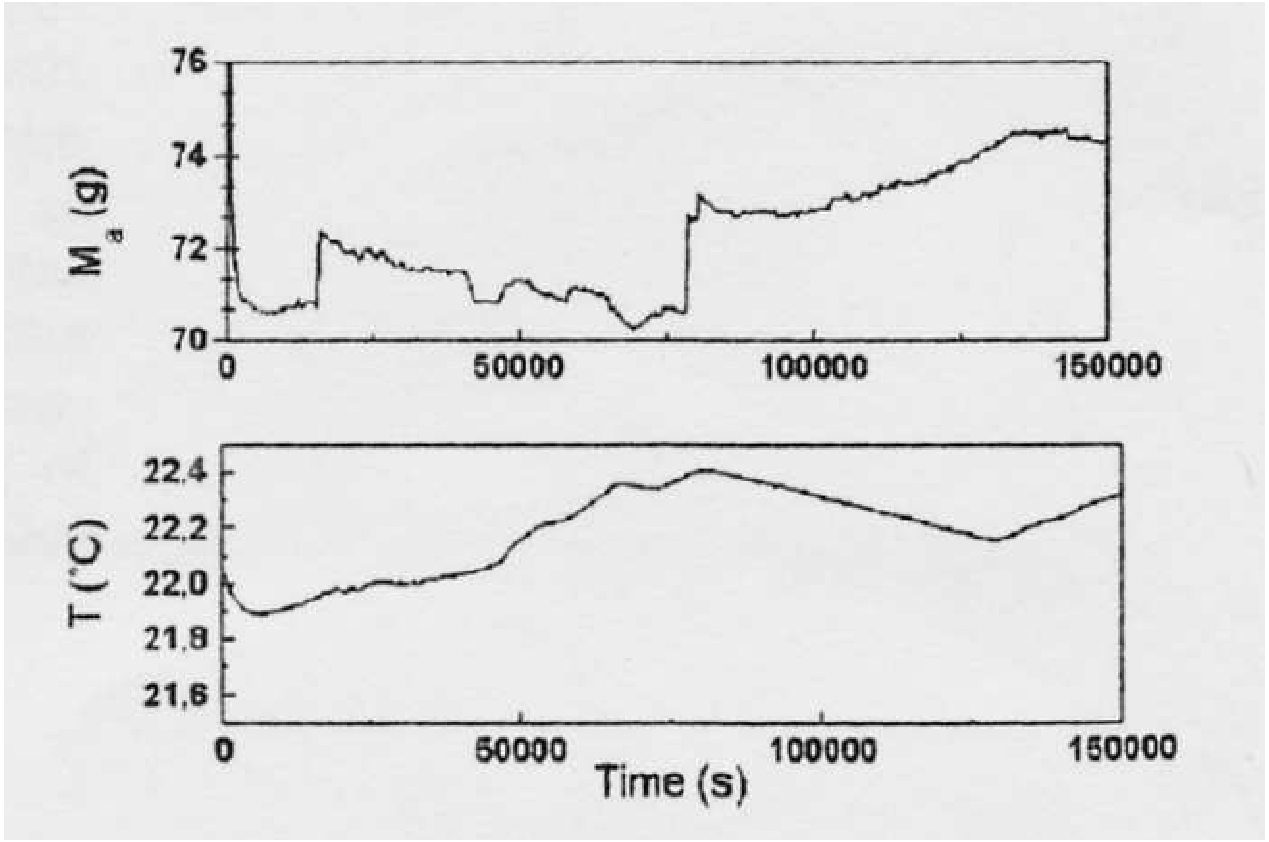}
\end{center}
\caption{Apparent mass and temperature variations. The system (see text) consists in a vertical cylinder (radius 2.0~cm) filled with glass beads (typical diameter 3~mm). A drift in the temperature of 0.4$^{\circ}$C leads to several reorganizations inside the pile and thus to several variations of the apparent mass of the system. Reprinted from \cite{Clement1997}.} \label{fig.Clement}
\end{figure}

\subsection{Apparent mass fluctuations}

In 1997, E. Cl\'ement and co-workers reproduced Jansen experiments \cite{Clement1997,Vanel1999}, which consists in measuring the apparent mass of a granular pile confined in a tube. A piston at the tube bottom, implemented on an electronic scale, makes it possible to measure the apparent mass of the pile. Although the experiment is in good agreement with Jansen's prediction \cite{Jansen1895}, they noticed that their data presented fluctuations as high as 20~\% in the saturated limit, where the measured mass becomes independent of the amount of beads poured inside the tube. Part of this fluctuations were attributed to temperature fluctuations and the authors emphasized that ``\textit{the origin is mixed since it can be due to the dilation of the boundaries and the resulting action on the piston or it can be due to the dilation of the grains, themselves inducing spontaneous rearrangements of the force network}" \cite{Clement1997}. In particular, the contribution of the boundary dilation will be addressed in section~IV. Figure~\ref{fig.Clement} illustrates that even a drift in the temperature of 0.4$^{\circ}$C is sufficient to trigger rearrangements which lead to measurable mass fluctuations. However, no systematic measurements were performed on this Jansen configuration and it would still be of great interest to assess the effect of temperature variations on apparent mass of a pile \cite{deGennes1999}.  

\begin{figure}[th]
\begin{center}
\includegraphics[width=0.45\textwidth]{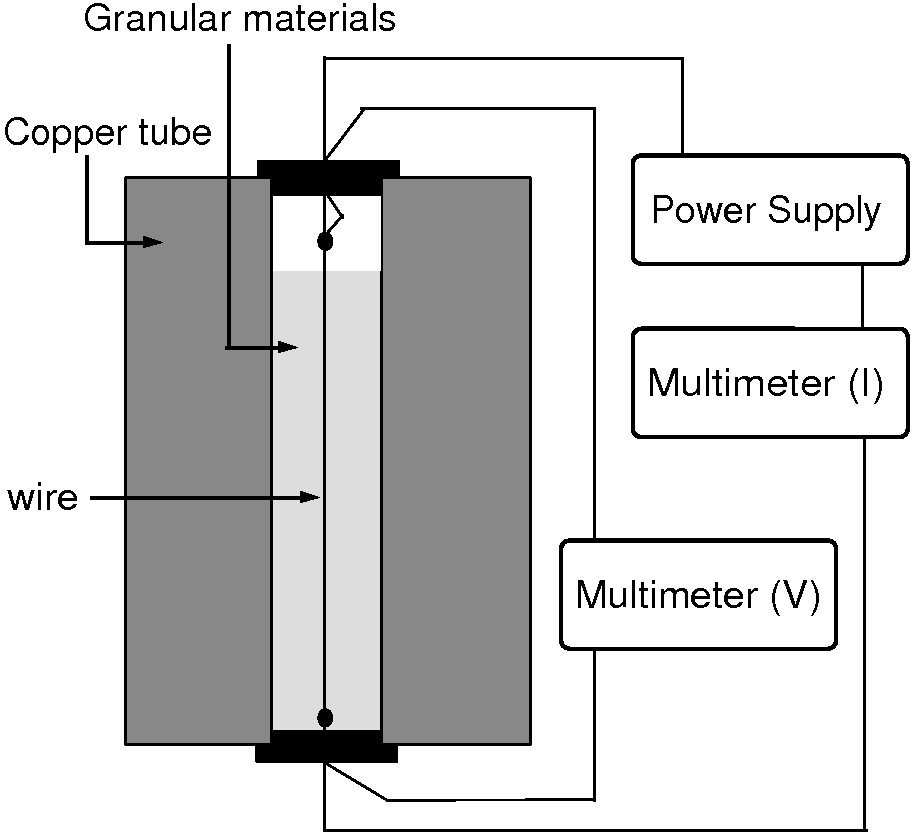}
\end{center}
\caption{Sketch of the experimental setup used to measure the effective thermal conductivity $\lambda_g$. The granular material consists in glass beads (typical diameter $300~\mu$m). The copper tube dimensions are the followings: inner diameter 1~cm; outer diameter 5~cm; length 15~cm. Reprinted from \cite{Divoux2009}.} \label{Setup}
\end{figure}

\section{... to controlled variations of temperature}
\label{control}

\begin{figure}[bh]
\begin{center}
\includegraphics[width=0.45\textwidth]{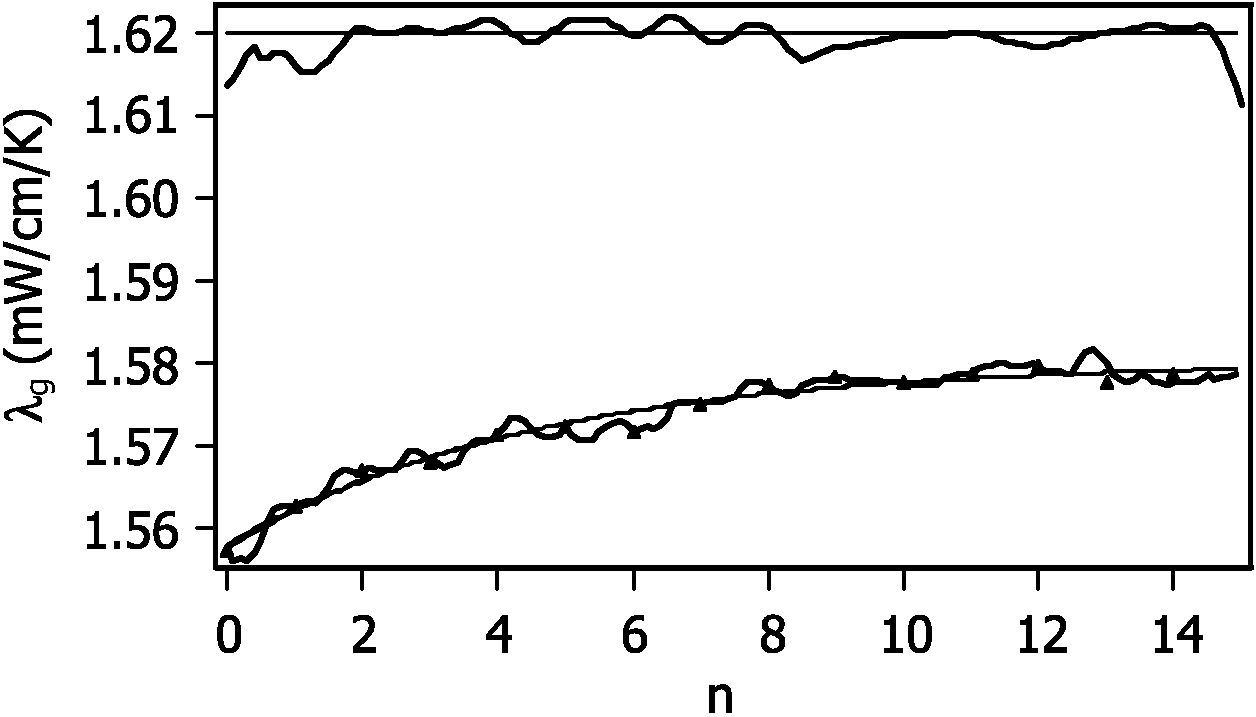}
\end{center}
\caption{Thermal conductivity $\lambda_g$ vs. number of cycles $n$. The upper curve corresponds to a pile which has been gently tapped prior to the experiment. The lower curve corresponds to a loose pile. Note that the conductivity is larger for a larger density. Moreover, the conductivity of the loose material increases significantly when one imposes the thermal cycles (typical grain diameter 300~$\mu$m, $\Delta T=40^{\circ}$C). Reprinted from \cite{Divoux2009}.} \label{fig.conductivite}.
\end{figure}


Since then, temperature variations have been used to generate minute perturbations inside a granular medium. Local heaters placed at different positions inside a static pile have been shown to produce very different disturbances on sound propagation underlining the key role of force chains in the propagation process \cite{Liu1994}. Moreover, small thermal perturbations have been used to generate large non-Gaussian conductance fluctuations in a 2D packing of metallic beads. Perturbations are induced by a 75~W light bulb standing a few centimeters away from the packing and experiments were performed with light turned on for packing prepared with light off, and vice versa. Gathered in bursts, such conductance fluctuations were interpreted as the signature of individual bead creep rather than collective vault reorganizations \cite{Bonamy2000,Bonamy2001}. In particular, the probability density function $P(\Delta t)$ where $\Delta t$ denotes the waiting time between two successive bursts of conductivity, follows a power law:
\[ P(\Delta t) \sim \Delta t^{-(1+\alpha_t)}\]
The exponent for stainless steal beads is found to be $\alpha_t \simeq 0.6$ independent of both the strength of the perturbation and of the external stress. The value of $\alpha_t$ is only governed by the surface roughness of the beads which has been confirmed by AFM measurements of the Hurst exponent of the bead surface \cite{Bonamy2000}. In agreement with the SAM \cite{Claudin1997} discussed in section~I, the origin of the fluctuations of conductivity originate in local micro-contact rearrangements.

More recently \cite{Divoux2009}, temperature variations have been applied by means of a nickel wire (diameter $r_w$~=100~$\mu$m) connected to a power supply, and which crosses a granular medium partially filling a copper tube (Fig.~\ref{Setup}). The tube is maintained at a constant temperature $T_{e}$ (precision $0.1^{\circ}$C). The imposed current $I$ and the resulting voltage $U$ are simultaneously measured in order to estimate the heating power $P = U I$ and the wire temperature $T_w$ which is deduced from the resistance $R_w = U/I$ (Fig.~\ref{Setup}). This setup makes it possible to measure the effective thermal conductivity $\lambda_g$ of the granular material as $T_w-T_e \propto P/\lambda_g$. 
One obtains $\lambda_g = 0.162 $~W/m/K which is compatible with the value expected for a pile of glass beads ($\lambda_{glass}=1.4$ W/m/K) surrounded by air ($\lambda_{air}=0.025$ W/m/K) \cite{Geminard2001}.
More interestingly, the authors also observe that for the same material $\lambda_g$ depends on the preparation~: One measures $\lambda_g \simeq 0.162$ W/m/K if the system is tapped prior to the measurement and $\lambda_g \simeq 0.156$ W/m/K if not. It is then particularly interesting to consider the behavior of the sample when subjected to several temperature cycles. When the measurements are repeated several times, one observes that the thermal conductivity of the loose sample along with its packing fraction significantly increase with the number $n$ of imposed cycles. By contrast, the conductivity of the tapped sample only slightly fluctuates around a constant value (Fig.~\ref{fig.conductivite}). Such results clearly emphasize that one can induce compaction by thermal cycling. This issue is addressed in the following section.

\begin{figure}[t]
\begin{center}
\includegraphics[width=0.4\textwidth]{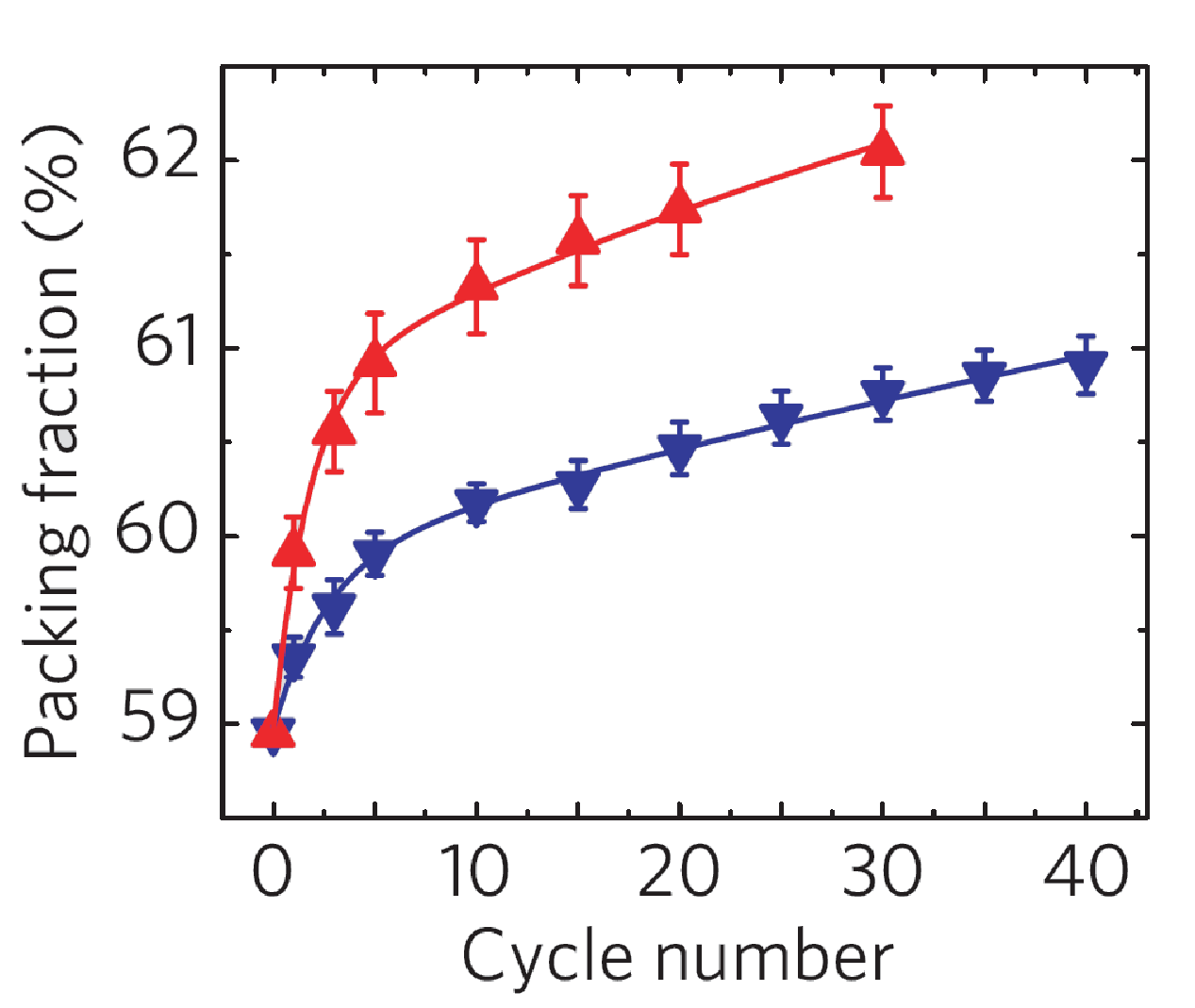}
\end{center}
\caption{Change in packing fraction after multiple thermal cycles (cycle temperatures: red, $\Delta T=107\pm 2^{\circ}$C; blue, $\Delta T=41\pm 1^{\circ}$C) for glass spheres in a plastic cylinder. Lines represent fits of the data by the sum of two exponentials introducing a short (resp. long) timescale associated to individual (resp. collective) rearrangements of the grains \cite{Mehta1991,Barker1993}. Reprinted with permission from \cite{Chen2006}. Copyright (2006) by the Nature Publishing Group.} \label{Chen}
\end{figure}

\section{Compaction induced by thermal cycling}
\label{cycles}

The compaction of a granular pile has been studied in two different configurations: (i) both the grains and the container are submitted to cycles of temperature \cite{Chen2006,Divoux2008,Chen2009}, and (ii) the grains alone experience the cycles \cite{Geminard2003,Divoux2009b}. In both cases, the packing fraction increases under thermal cycling. We first review the role of the amplitude and frequency of the cycles for both cases. Second, we gather some insights on the dynamics at the grain scale.

\subsection{The role of the cycling amplitude and frequency}

Chen and co-workers \cite{Chen2006,Chen2009} examined the change in packing fraction for glass, polystyrene or high density polyethylene spheres (typical diameter 1~mm) contained in polymethylpentene plastic or borosilicate glass cylinders (diameter ranging from 14 to 102~mm) in response to thermal cycling. One thermal cycle is conducted by placing both the grains and the container into an oven until the thermal equilibrium is reached, before letting them relax at ambient temperature. The typical cycling period is about 10 hrs and has not been varied. They observed that the packing fraction increases, even after one cycle \cite{Chen2006}. This effect is independent of the sample depth and width (14-102~mm), more efficient for higher cycling amplitudes (Fig.~\ref{Chen}) and, contrary to their first guess, also independent of the relative coefficients of thermal expansion of the grains and the container \cite{Chen2009}, as confirmed by numerical simulations \cite{Vargas2007}. This indicates that the compaction mechanism is local as suggested by the observations discussed in section~III and thus linked to the solid friction between the beads, and between the beads and the container. 
\begin{figure}[t]
\begin{center}
\includegraphics[width=0.45\textwidth]{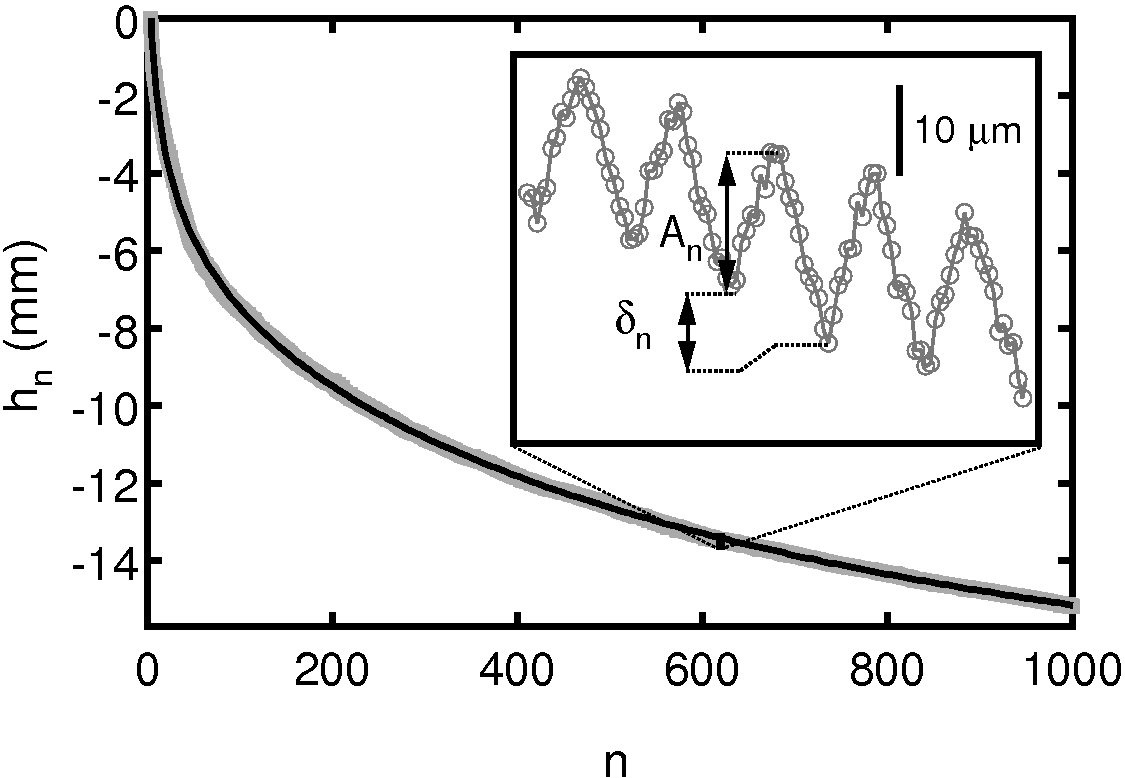}
\end{center}
\caption{Height variation $h_n$ vs. number of cycles $n$. One observes first an exponential behavior at short time followed by a logarithmic creep at long time. The black curve corresponds to the test function $h_n^t\equiv\,h_0+h_e\,\exp{(-n/n_c)} + h_l\,\ln(n)$. Inset - Oscillations of the column height associated with the temperature cycles~: $A_n$ and $\delta _n$ are respectively defined to be the amplitude of the increase and the drift of $h_n$ at the cycle $n$ ($H = 140$~cm, $2 \pi/\omega = 600$~s and $\Delta T=10.8^{\circ}$C.). Reprinted figure with permission from \cite{Divoux2008}. Copyright (2008) by the American Physical Society.} \label{Hauteur}
\end{figure}

Two other setups have been used to impose cycles of temperature: the first one consists of a vertical glass tube filled with spherical glass beads \cite{Divoux2009,Divoux2008}. The temperature cycles are imposed by means of a heating cable (40~W/m) directly taped on the outer surface of the tube wall. Here again both the container and the grains are thus submitted to the cycles. The free surface of the material is imaged from the side with a video camera which makes it possible to measure accurately the column height $H$ from the images and to perform a \textit{time resolved study} of its dynamic. In such a setup one can control both the amplitude $\Delta T$ of the cycles, and the penetration length $l_p\equiv \sqrt{ 2\lambda/(\rho\, C \omega)}$ through the frequency $\omega /2\pi$ of the cycles ($\lambda$ and $C$ respectively denote the thermal conductivity and heat capacity of a typical glass-grains pile).
Prior to each experiment, the granular column is prepared in a low-density state thanks to a dry-nitrogen upward flow. The top of the column is then higher than the field imaged by the camera and one sets the amplitude of the cycles, $\Delta T$, to the largest accessible value, $\Delta T=27.1^{\circ}$ C. The column flows under thermal cycling, and the preparation of the sample ends when the top of the column enters the observation field. At this point, the granular column is ``quenched": the amplitude of the cycles is set to the chosen value $\Delta T$ lying between $0$ and $27.1 ^{\circ}$C, which defines the origin of time $t= 0$. The granular column is subsequently submitted to at least $1000$ cycles.
\begin{figure}
\begin{center}
\includegraphics[width=0.45\textwidth]{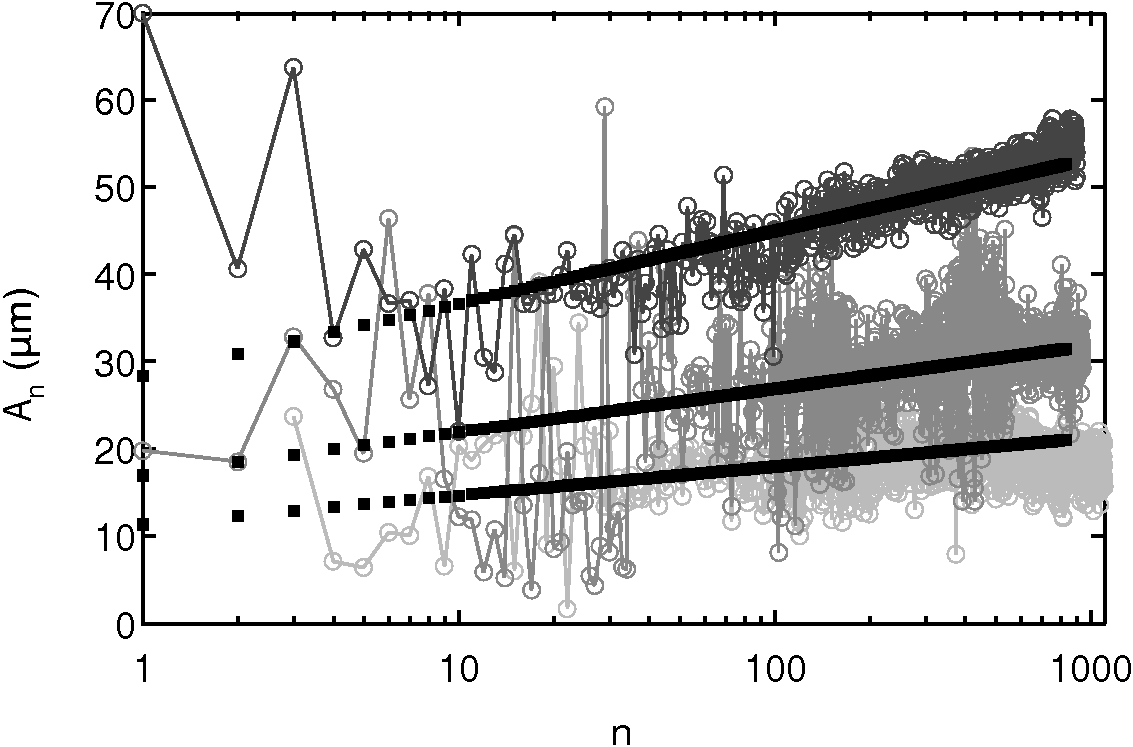}
\end{center}
\caption{Amplitude $A_n$ (defined in Fig.~\ref{Hauteur}) versus the number of cycles n. The data are successfully accounted for by $A_n=\Delta T\,[a_0+b_0\ln(n)]$ with $a_0=1.0$~$\mu$m$\cdot$K$^{-1}$ and $a_0=0.13$~$\mu$m$\cdot$K$^{-1}$ (initial column height: 140~cm, cycling period: $2\pi/\omega=600$~s and, from top to bottom, $\Delta T$=27.1, 16.2 and 10.8$^\circ$C). Such an increase in the dilation amplitude with the cycles of temperature is a signature of the aging the granular pile experiences under thermal cycling \cite{Divoux2009}. Reprinted figure with permission from \cite{Divoux2008}. Copyright (2008) by the American Physical Society.} \label{AN}
\end{figure}

\textit{Role of cycling amplitude}-  Here, the cycling period (10 minutes) is chosen to ensure that the associated  thermal penetration length $l_p \simeq 6$~mm is about the tube radius so that all the grains experience the dilation process. Figure~\ref{Hauteur} reports the variation of the column height defined as $h_n \equiv  H(2\pi n/\omega)-H(0)$, where $n$ denotes the number of imposed cycles. For a large $\Delta T$ (typically more than 3$^{\circ}$C), the column systematically compacts (Fig.~\ref{Hauteur}) at each cycle [Fig.~\ref{Hauteur},~(inset)]. Following Barker and Mehta \cite{Mehta1991,Barker1993} like Chen and co-workers \cite{Chen2006,Chen2009}, one can try to fit the height decay by the sum of two exponentials: the result is not satisfying and the compaction dynamics is better accounted for by the following test function: $h_n^t\equiv\,h_0+h_e\,\exp{(-n/n_c)} + h_l\,\ln(n)$ (Fig.~\ref{Hauteur}) \cite{Divoux2009b}. First, this response to the thermal quenching is strikingly similar to the one the system exhibits to vanishing step strain perturbations \cite{Brujic2005}. Second, the long time logarithmic behavior of the column height leads to an inverse logarithmic evolution for the packing fraction $\phi_n$
which is strongly reminiscent of the way the packing fraction of the granular pile submitted to tapping evolves in the absence of any convection \cite{Knight1995}. Another remarkable feature of this time resolved study is that one can also observe that $A_n$, defined as the amplitude of the increase of $h_n$ at the $n$-th cycle (Fig.~\ref{Hauteur}), is proportional to the cycling amplitude $\Delta T$ and increases logarithmically with $n$ (Fig.~\ref{AN}). The more the column compacts, the higher is the dilation amplitude of the total height of the column at each cycle. This is a signature of the aging of the pile under thermal cycling, like the increase of the thermal conductivity (section~III).\\ 
In the limit of small cycling amplitude $\Delta T$ (here below $\Delta T_c=3^{\circ}$C), one observes that the column is not flowing regularly anymore, but evolves by successive collapses (typical amplitude of a tenth of grain diameter) separated by rest periods (randomly distributed) [Fig.~\ref{Hauteurlog}~(top)]. Such a different compaction process from the one observed for high cycling amplitude has been linked to the surface roughness of the glass beads. Indeed, for $\Delta T<\Delta T_c$ the dilation of a grain is smaller than the surface roughness, and thus, the beads behave as smooth particles whose dilation induces only localized rearrangements. Thus one has to wait several cycles of temperature before observing a large scale collapse of the column level, as a cumulative effect of several local reorganizations. On the contrary, for $\Delta T>\Delta T_c$ the dilation of a grain is larger than the surface roughness, and thus the beads behave as rough particles. One cycle of temperature is more likely to generate a large scale reorganization \cite{Divoux2008}.

\begin{figure}[th]
\begin{center}
\includegraphics[width=0.45\textwidth]{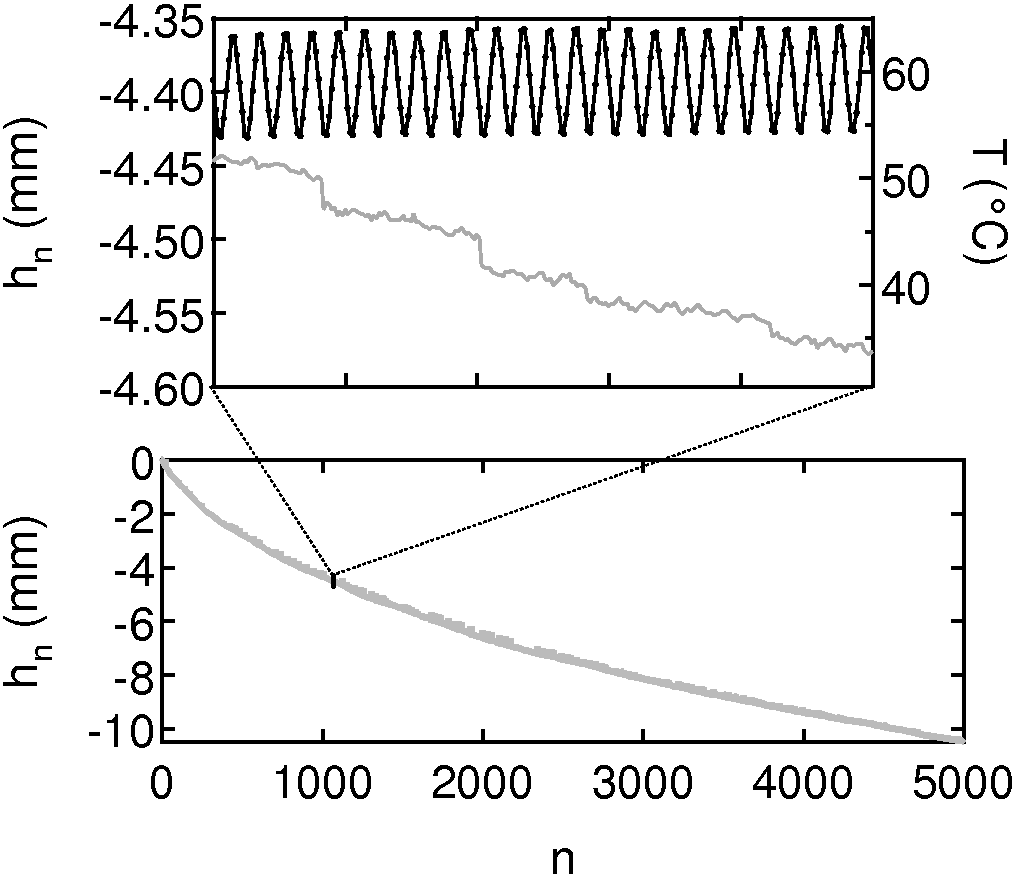}
\end{center}
\caption{Evolution of the column height $h_n$ versus the number $n$ of cycles of temperature. Inset: zoom on the evolution of $h_n$ over 24 cycles of temperature: the column settles by jumps separated by periods of continuous flowing  ($H=140$~cm, $T=2\pi / \omega=150$~s et $\Delta T=9.5^{\circ}$C). Reprinted from \cite{Divoux2009b}.} \label{fig.freq}
\end{figure}

\begin{figure}[th]
\begin{center}
\includegraphics[width=0.45\textwidth]{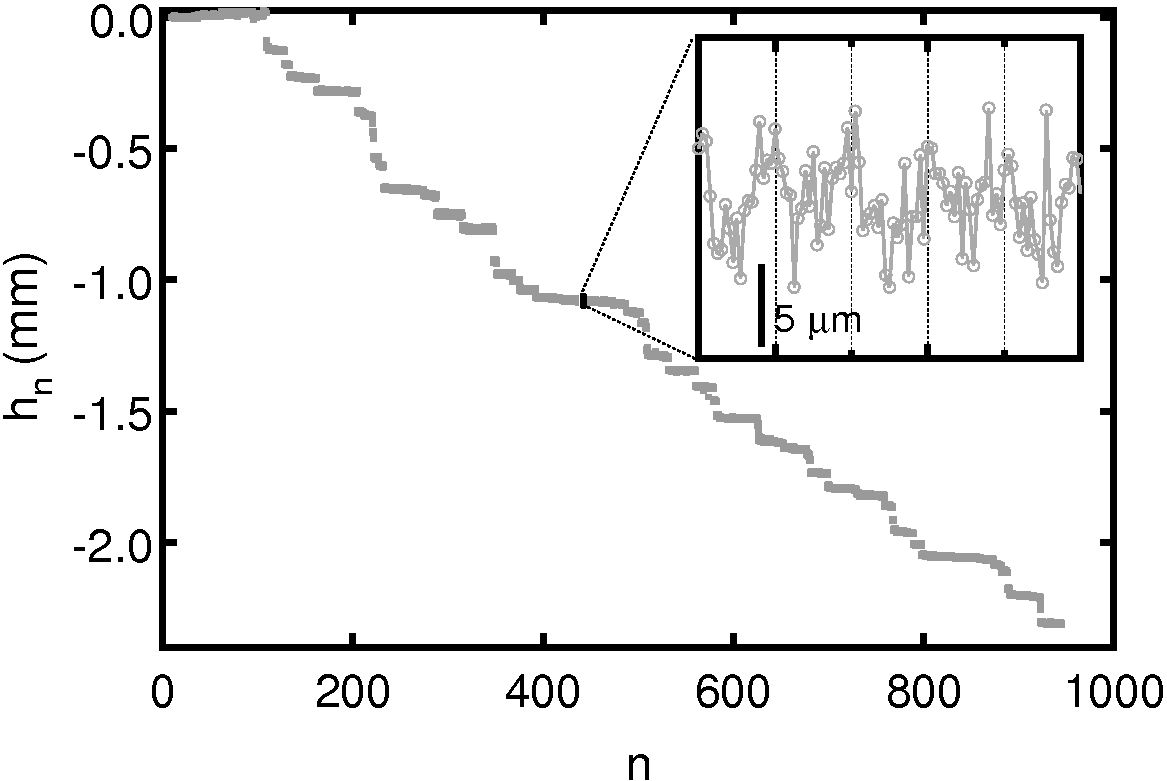}
\end{center}
\begin{center}
\includegraphics[width=0.45\textwidth]{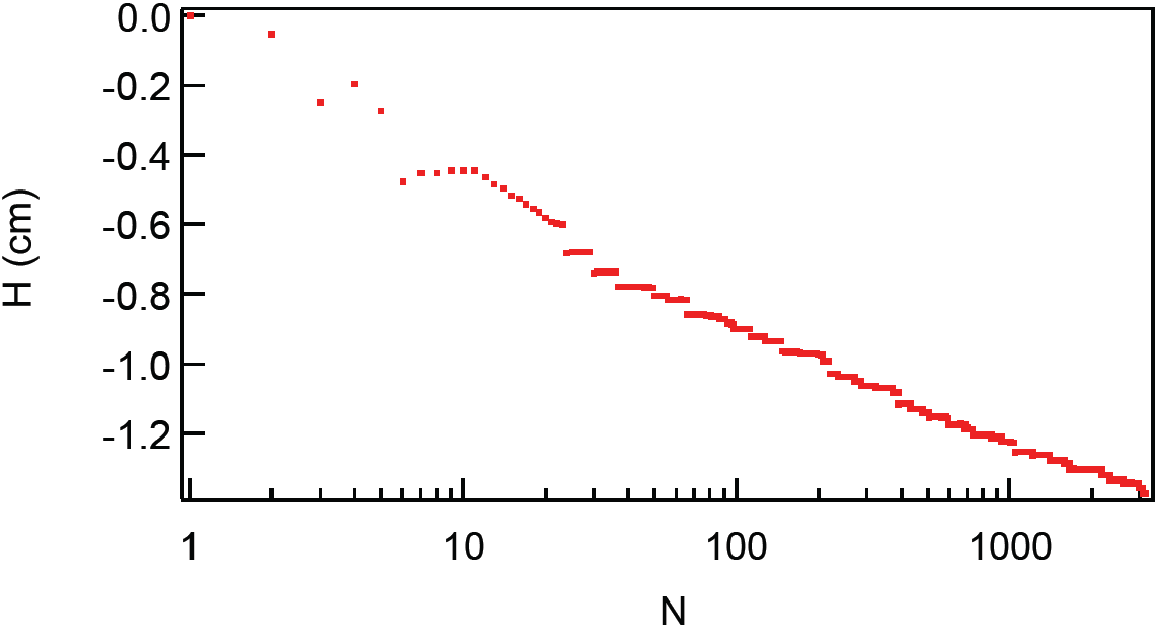}
\end{center}
\caption{Height variation $h_n$ of the column versus the number of cycles of temperature. (Top) Curve obtained in the case where temperature cycles were applied to both the container and the grains. On average, $h_n$ decreases linearly with the number of cycles and the column settles by jumps separated by rest periods ($H=140$~cm, $2\pi/\omega=600$~s, and $\Delta T=$2.8$^{\circ}$C). Reprinted figure with permission from \cite{Divoux2008}. Copyright (2008) by the American Physical Society. (Bottom) Curve obtained in the case where cycles of temperature are applied to the grains alone by means of a hot wire crossing the pile (see Fig.~\ref{Setup}). The column also settles by jumps, but the height variations goes as the logarithm of the number of cycles. Reprinted from \cite{Geminard2003}.} \label{Hauteurlog}
\end{figure}

\textit{Role of cycling frequency}- The influence of the frequency is discussed in \cite{Divoux2009b}, but still remains to be fully explored. On the one hand, for frequencies shorter than the one discussed above, the penetration length is larger than the tube radius. The compaction process is more efficient, and the dilation amplitude $A_n$ is constant, independent of the age of the system. On the other hand, for larger frequencies, the penetration length is shorter than the tube radius. The column is observed to flow continuously while small amplitude settlings may occur (Fig.~\ref{fig.freq}). The interpretation proposed in \cite{Divoux2009b} is the following: the penetration length is shorter than the tube radius, the grains in the center of the column are not submitted to the cycles of temperature and thus behave as a solid body that experiences some stick-slip motion due to the periodic dilations of the grains close from the walls of the container. Indeed, the column dynamics over a few cycles [Fig.~\ref{fig.freq}, (inset)] displays a very similar behavior to a tilted monolayer of grains on an inclined plane (see Fig.~3 in \cite{Scheller2006}). In this case, the flow results from a competition between a gravitational flow and the formation of arches \cite{Scheller2006}. However, a full study on the role of the frequency, and in particular on the influence of harmonics in the shape of the cycles (squares, tooth-like signals, etc.) on the compaction dynamics still remains to be done.

The second setup that has been used to impose temperature cycles is very similar to the one previously discussed in figure~\ref{Setup}. A thin wire crosses a granular column in its center, while this time the container is a thin glass tube \cite{Geminard2003}. For large enough frequencies, the penetration length is shorter than the tube radius which makes it possible to cycle the grains close to the wire and let the container at rest. It is here relevant to compare the way the dilation of the container modifies the compaction process in the limit of low amplitude cycles. In the case where both the grains and the container are submitted to variations of temperature, the compaction process is linear in the number $n$ of applied cycles [Fig.~\ref{Hauteurlog} (top)], whereas, in the case where the grains alone are submitted to cycles of temperature and the container is fixed, the compaction process goes logarithmically with $n$ [Fig.~\ref{Hauteurlog}~(bottom)]. Both experiments were conducted under the same cycling frequency, and for comparable cycling amplitudes (resp. $\Delta T$ = 2.8 and 1$^{\circ}$C). Those results clearly indicate that the dilation of the container plays a key role in the global compaction dynamics of the granular column. However, up to now, the full contribution of the container dilation has not been characterized. The answer should be easily obtained with the set-up presented in figure~\ref{Setup}.  

\subsection{Dynamics at the grain scale under thermal cycling}

\begin{figure}[t]
\begin{center}
\includegraphics[width=0.45\textwidth]{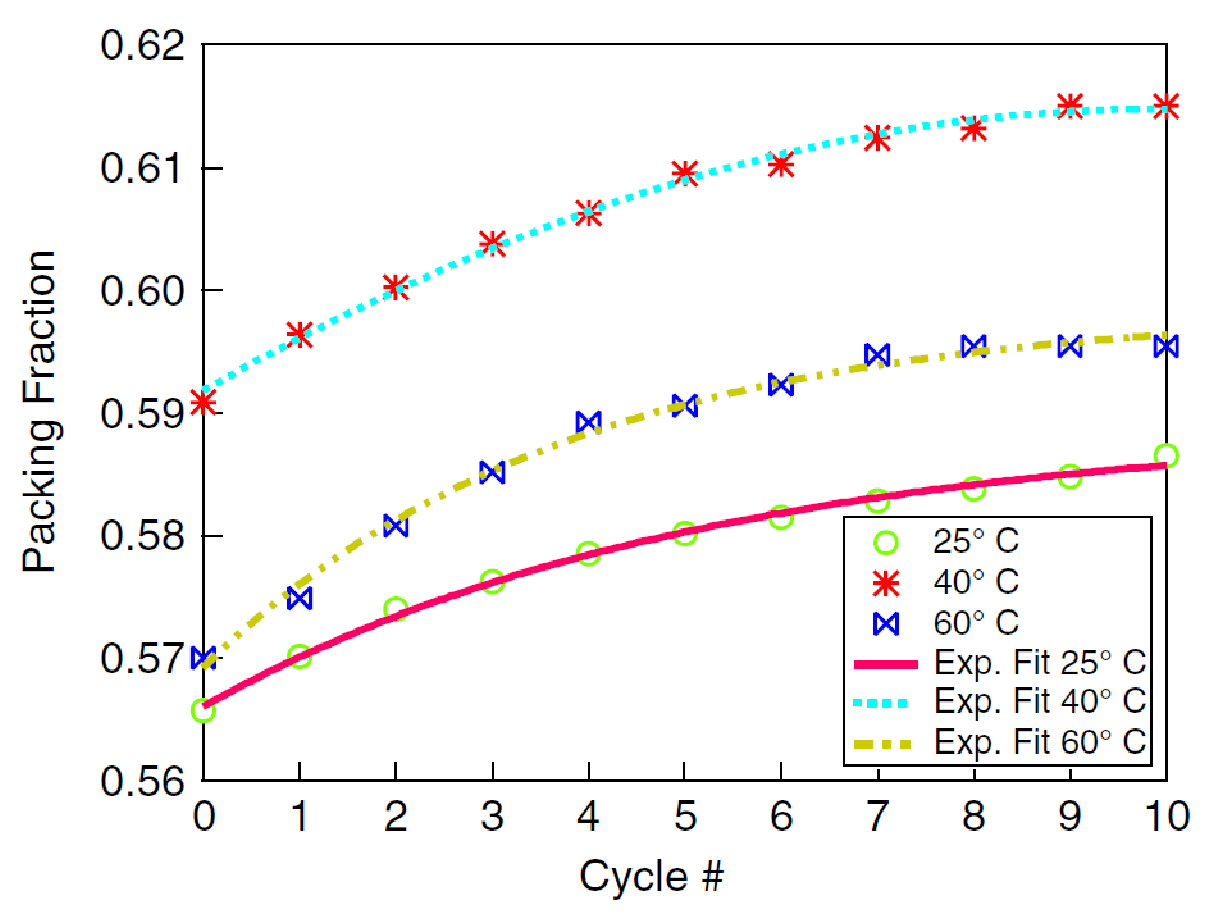}
\end{center}
\caption{Packing fraction vs the number of cycles of temperature for various temperature differentials. The system consisting of glass beads immersed in an index matching oil inside a cylinder, is subjected to thermal cycling via a water bath. A weight is placed on top of the bead packing to apply a controlled vertical force. Both this weight and the interstitial fluid might explain the rapid compaction compared to what is observed in Fig.~\ref{Hauteur}. Reprinted figure with permission from \cite{Slotterback2008}. Copyright (2008) by the American Physical Society.} \label{fig.losert}
\end{figure}

One of the key features of the motion induced by thermal cycling in a granular pile is that a grain stays in contact with its neighbors and that the global motion is really slow (hours and days typically). It makes it possible to use experimental techniques developed to study creep motion, like dynamic light scattering (DLS) \cite{Crassous2008}, successive snapshots \cite{Divoux2008} or 3D scanning \cite{Slotterback2008} to follow the individual grain trajectories. Recently, L. Djaoui and J. Crassous used a DLS setup to follow the evolution of a granular pile under thermal cycling \cite{Djaoui2005}. This method was successful at extracting both linear and non-affine displacement of the grains \cite{Crassous2009}, as recently done under mechanical shear \cite{Utter2008}, thus confirming that controlled temperature variations are a delicate way to generate small displacements.
Slotterback and co-workers have been following the dynamics induced by thermal cycling at the grain scale \cite{Slotterback2008}. In their experiment, the column of grains is immersed in an index-matching oil containing a laser dye which makes it possible to produce 3-D images of the system at the end of each cycle by means of a laser sheet scanning method. Particle displacements in this jammed fluid correlate strongly with rearrangements of the Voronoi cells defining the local environment about the particles. This might be a proof for stringlike cooperative motion as already pointed out in quasi-2D air driven granular flows \cite{Keys2007}, and more generally in supercooled liquids \cite{Donati1998,Douglas2006}.
However, it is rather difficult to compare these results to those obtain for dry grains \cite{Chen2006,Divoux2008}. First, because in the case of the immersed experiment, a weight is placed on top of the granular column to apply a controlled vertical force, whereas in the dry case the top column is free of any weight. Second, the presence of the interstitial fluid lubricates the contacts between the grains, and certainly leads to a more homogeneous propagation of temperature front through the pile \cite{Vargas2002}. Indeed, the thermal conductivity of oil is roughly 10 times higher than the thermal conductivity of air which reduces the role played by force chains in the heat conduction. Those three effects speed up the compaction process compared to the dry case: the steady state packing fraction is obtained after only 10 cycles in the immersed system (Fig.~\ref{fig.losert}).

\subsection{A comparison with the tapping experiments}

 Let us here recall that {\it tapping experiments} consist in imposing vertical vibrations to a container full of grains. In practical terms, a low frequency signal (usually $10<f=\omega/2\pi<100$~Hz) \cite{Remark2} is used to generate a ``tap" of amplitude $a$, two successive taps being separated by a duration $\Delta t$ long enough to be considered as independent (roughly $\Delta t \simeq 1$~s). The key control parameter is the reduced acceleration $\Gamma$ defined as $\Gamma \equiv a\omega^2/g$ \cite{Remark1} and one can distinguish between three different regimes. For low values of the reduced acceleration ($0\leq  \Gamma\leq  \Gamma^*\simeq 1.2$), where $\Gamma^*$ denotes the critical acceleration at which the grains lose contact with the bottom of the container, the compaction process is extremely slow \cite{Philippe2002b}. This regime is very similar to that observed for thermal cycling: the geometry of the pile is essentially frozen and the force network can still evolve by slowly depleting the most fragile contacts \cite{Kabla2004,Umbanhowar2005}. Besides, under such small vibrations, the packing is aging: the behavior of the pile is first dominated by grain motion and then by the contact force variations at larger timescales \cite{Umbanhowar2005}. However, in this range of acceleration, no phenomenological law was proposed to describe the evolution of the packing fraction and/or the column height that could be compared to the results obtained under thermal cycling. For intermediate values of the acceleration ($\Gamma^*\simeq 1.2 \leq  \Gamma \leq  \Gamma_c \simeq 2$), the compaction process is more efficient and takes place over timescales accessible in the lab \cite{Knight1995,Philippe2002b}. At each tap, the packing loses contact with the bottom of the container, and crushes back which generates a shockwave responsible for the compaction of the pile. A convection roll might take place inside the container - roughly, if the ratio of the container size to the typical bead diameter is large enough, a signature of which can be found on the free surface of the pile that presents a slope \cite{Evesque1989}. Such convection rolls are not observed in a pile submitted to cycles of temperature as confirmed by the results observed under tapping for higher range of acceleration. 
Indeed, for larger values of the acceleration ($\Gamma > \Gamma_c \simeq 2$), the compaction takes place homogeneously in the whole shaken sample \cite{Knight1995,Philippe2002}, but the dynamics strongly depends on the presence or the absence of convection rolls inside the pile \cite{Richard2005,Philippe2002,Ribiere2005c}. Without any convection roll, the packing fraction $\phi_n$ evolves in a similar way to what was found for thermal cycling in \cite{Divoux2008}:
\begin{equation}
\phi_{\infty}-\phi_n \propto \frac{1}{\ln (n)} \label{eq1}
\end{equation}
where $\phi_{\infty}$ is the steady state packing fraction obtained at long time. This phenomenological expression, first proposed in \cite{Knight1995}, has been justified theoretically using various approaches: analogy to a parking process \cite{Krapivsky1993,BenNaim1998}, ``tetris-like" models \cite{Caglioti1997,Nicodemi1997}, excluded volume approach \cite{Boutreux1997} and void diffusion models \cite{Linz1996,Linz1999}. A common feature of these models is the geometrical frustration: the denser  the packing, the harder it becomes to insert grains from the top. Such a frustration process seems to be also at stake in the compaction induced by thermal cycling, which still remains to be properly put into equations. Still, under tapping, if convection rolls develop inside the pile, the compaction dynamics follows a Kohlraush-Williams-Watts (KWW) law:
\begin{equation} 
\phi_{\infty}-\phi_n \propto  \exp \left[-(n/n_f)^{\beta} \right] 
\end{equation}
where $n_f$ and $\beta$ are two parameters which depend on the acceleration $\Gamma$ \cite{Philippe2002b,Philippe2002,Remark3}. Such a law pops up naturally if one considers the global dynamics at stake as a superposition of several processes, each with a proper and well defined characteristic time. Indeed, one can see here the compaction process as the sum of the reorganization of several groups of beads, each with different sizes and different relaxation times. One can also emphasize that the global dynamics is not controlled by the biggest (and thus slowest) groups of beads, but by the fastest individual beads. These are the grains that can quickly relax individually and jump over large distances (roughly their radius), and so during a single tap, that control the compaction dynamics \cite{Ribiere2005}. This result indicates once again that there are certainly no convection rolls inside a granular pile submitted to cycles of temperature as neither the KWW law nor rare jumps of individual beads are observed experimentally in this case \cite{Remark4}.



\begin{figure}[th]
\begin{center}
\includegraphics[width=0.45\textwidth]{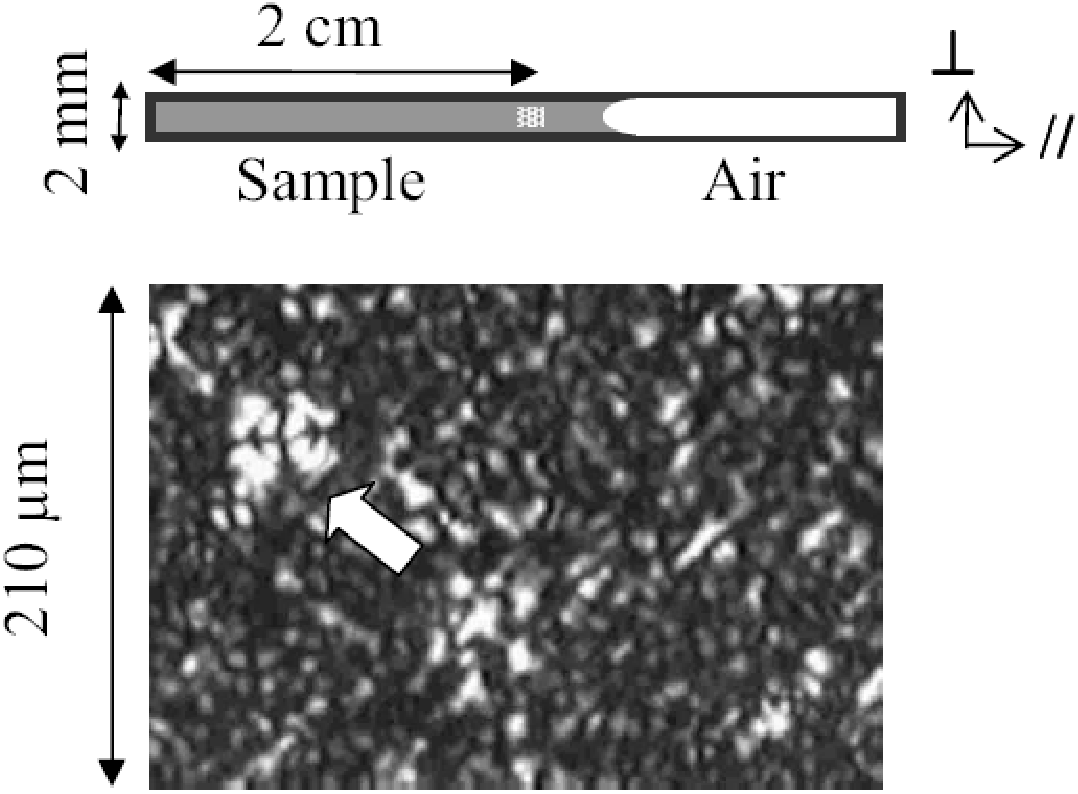}
\end{center}
\caption{(Top) Scheme of the capillary containing the onion gel, the location of the field of view and the orientation of the parallel and perpendicular axis. (Bottom) Portion of a typical image of the sample taken by light microscopy between cross polarizers; The arrow points to a large onion. Reprinted figure with permission from \cite{Mazoyer2009}. Copyright (2009) by the American Physical Society.} \label{fig.Mazoyer_Setup}
\end{figure}

\section{Enlarged ``granular" systems}
\label{Enlarged}

In this section, we tackle two other systems different from a simple dry granular assembly and for which  temperature cycles might play a crucial role in the observed dynamics.

\subsection{Thermally driven aging in a polydisperse vesicle assembly}

\begin{figure}[th]
\begin{center}
\includegraphics[width=0.45\textwidth]{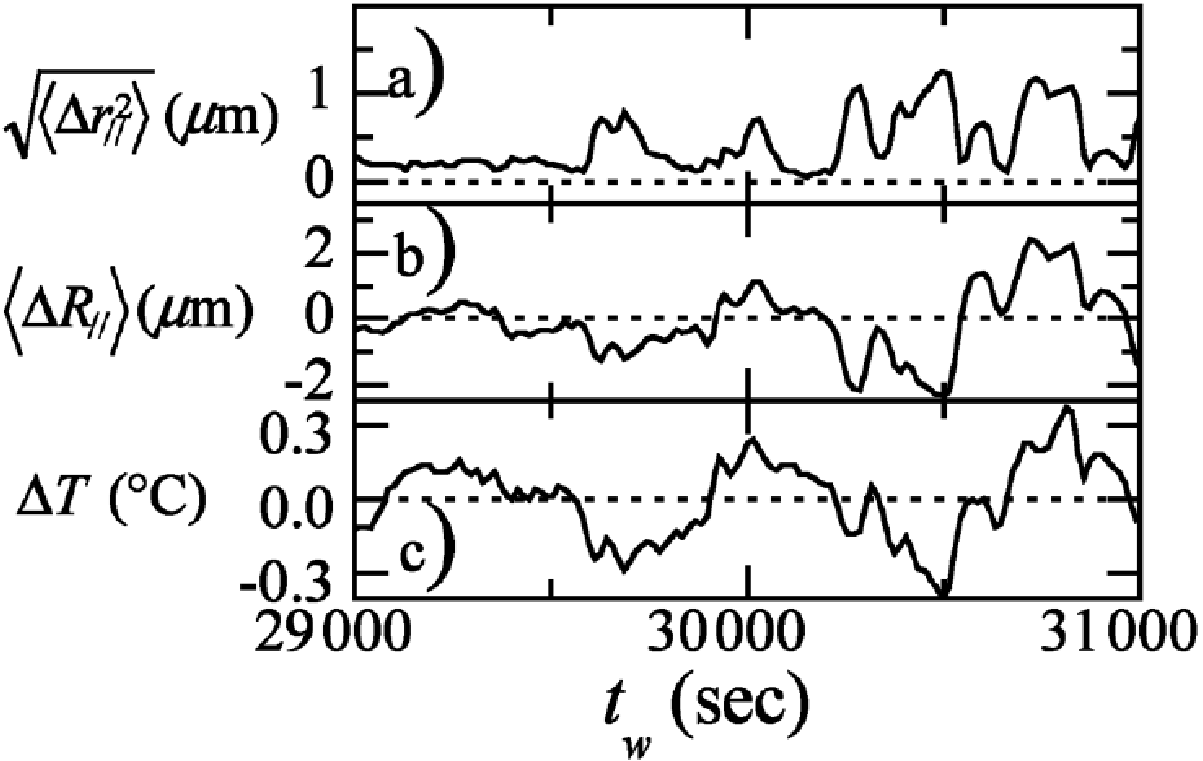}
\end{center}
\caption{Age dependence of (a) the square root relative parallel displacement, (b) the global parallel displacement and (c) the temperature variation $\Delta T$ over a lag time $\tau$. Reprinted figure with permission from \cite{Mazoyer2006}. Copyright (2006) by the American Physical Society.} \label{fig.Mazoyer}
\end{figure}

The closest case from what we have been presenting in this article consists in a dense packing of polydisperse multilamellar vesicles, or ``onions", submitted to (unavoidable) temperature fluctuations \cite{Mazoyer2006,Mazoyer2009}. Such a system, a water-based mixture of surfactants and block-copolymer, is fluid below 8$^{\circ}$C but forms a phase of jammed vesicles at ambient temperature (here 23.4$^{\circ}$C) which is known to experience aging \cite{Ramos2005,Ramos2001}. The system is loaded in a glass capillary which is flamed sealed and placed under a microscope. Images are taken every 15 sec during 24~h (Fig.~\ref{fig.Mazoyer_Setup}). The temperature is controlled within 0.09$^{\circ}$C. Mazoyer and co-workers observed that the unavoidable temperature fluctuations induce local mechanical shears in the whole sample due to local thermal expansion and contraction, which is a scenario very similar to the one proposed in \cite{Divoux2008} for dry grains. Here, each image [Fig.~\ref{fig.Mazoyer_Setup} (Bottom)] is divided into small regions of interest (ROI) and the authors look at the translation motion $\Delta R (t_w,\tau)$ of each ROI for pairs of images taken at two different times ($t_w$ and $t_w+\tau$). Both the global parallel displacement $\langle \Delta R_{//} \rangle (t_w)$ [Fig.~\ref{fig.Mazoyer} (b)] and the relative parallel displacement $[\langle \Delta r_{//}^2 \rangle]^{1/2} (t_w)$ [Fig.~\ref{fig.Mazoyer} (a)] present strong correlations with the temperature fluctuations $\Delta T$ [Fig.~\ref{fig.Mazoyer} (c)] \cite{Mazoyer2006}. Furthermore, looking at individual trajectories of ROIs, the authors identify two kinds of dynamical events: reversible and irreversible rearrangements. The first class is due to the contraction-elongation of the sample because of temperature fluctuations, and correspond to a shear deformation along the long axis of the capillary. The second class of events that are irreversible occurs as the result of repeated shear cycles. The motion resulting from these ultraslow rearrangements is ballistic (the initial growth of $\Delta r_{//}$ is proportional to $\tau$), with a velocity that decreases exponentially as the sample ages. Such results could be easily checked in dry and immersed granular systems \cite{Divoux2008,Slotterback2008} to see how far the analogy between a jammed vesicle assembly and a granular pile goes. Also, it raises the generality of such dynamics governed by temperature fluctuations. In the case of dilute colloidal systems \cite{Duri2006}, this scenario does not hold, as no correlation could be observed between temperature fluctuations and rearrangements. Indeed, one expects temperature fluctuations to play a key role in divided and athermal systems being rather concentrated or solid-like. It might therefore be relevant to find out other systems presenting temperature-induced strain fluctuations, in order to find how general the properties of these ultraslow rearrangements are.

\subsection{Stone heave phenomena}
 
\begin{figure}[th]
\begin{center}
\includegraphics[width=0.4\textwidth]{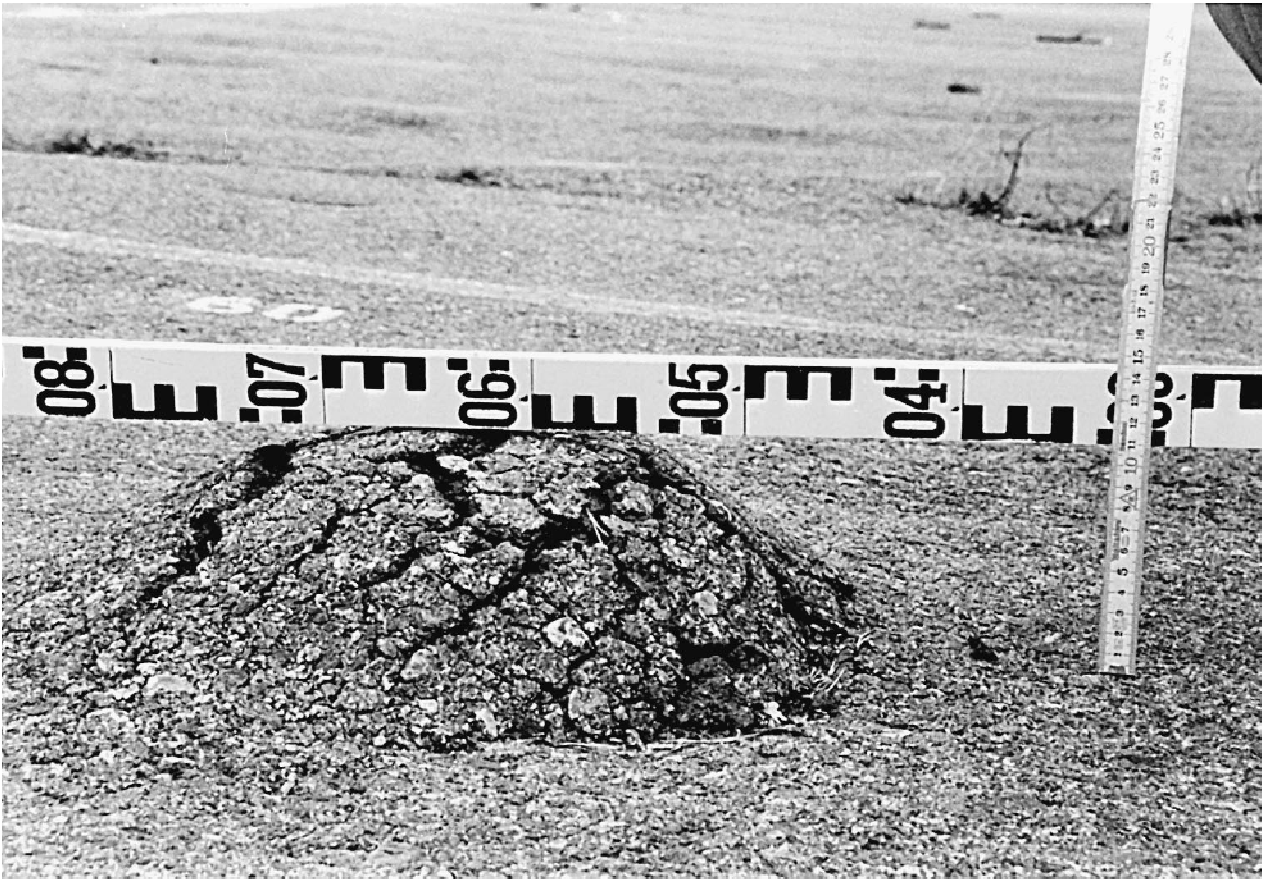}
\end{center}
\caption{Raised stone in a paved parking lot in Lule\aa{}, Sweden. Reprinted from \cite{Viklander2000}, Copyright (2000), with permission from Elsevier.} \label{fig.SH}
\end{figure}

Cyclic freezing and thawing of soils can cause stones and particles embedded in those soils to move and relocate mainly depending on the initial void ratio (defined as the fraction of sample filled by empty space). The stones move vertically upward (Fig.~\ref{fig.SH}) in dense soils (low void ratio) and vertically downward in loose soils (high void ratio). A full interpretation of this phenomenon is still lacking, but in the case of stone heave, the mechanism invoked lays on the propagation of the frost front in the soil: the freezing from the top leads to the growth of the ice beneath the stone, due to its higher thermal conductivity than that of the surrounding soil and in turn, causes the stone heave. A review of this topic can be found in \cite{Viklander1998}. A remarkable effect is that the void ratio also evolves toward a critical value (in the range of 0.29-0.34) under the successive freezing and thawing cycles \cite{Viklander1998}. Thus, a dense (loose resp.) soil will become looser (denser resp.) under cycles of freezing and thawing. This is exactly what experiences a granular assembly submitted to a mechanical shear: its packing fraction will tend to a critical value \cite{Aharonov1999}. Here again, as for onions and dry granular assemblies, the temperature-induced shear controls the dynamics. Such an analogy suggests that one could shed some new lights on the stone heave phenomenon by looking at the way a large intruder (simulating the stone) in a granular pile (simulating the soil) behaves under thermal cycling (which produces a shear \cite{Divoux2008}). Such a work has been tackled by Chen and co-workers in \cite{Chen2009}. They observed that, under thermal cycles, the intruder (aluminum or brass) either does not move, or sinks inside the pile (polystyrene beads in a borosilicate glass container) if the intruder density overcomes a certain threshold. The pressure exerted by the intruder on the pile seems to be the relevant parameter, despite the ratio of the container size to the grains diameter which might play a crucial role on the force network inside the pile, has not been varied. Also, no experiments have been performed varying the initial packing fraction or the initial position of the intruder in the granular packing to mimic the observations performed on stone heave \cite{Viklander1998,Kolstrup2010}. Still, such a simple setup is certainly relevant to extract the key ingredients of the stone heave phenomena, and deserves further use. A comparison of the results with those obtained for vibrated grains (Brazil nut effect \cite{Kudrolli2004,Ulrich2007}) would also be interesting. 

\section{Summary, open questions and outlooks}
\label{Conclusion}

\subsection{Summary}

Temperature variations, even of low amplitude, induce the reorganization of a granular assembly and its slow compaction, and lead to the aging of the system, a signature of which can be found in the slowing down of the dynamics and in the evolution of the key properties of the system (increase of the effective thermal expansion and thermal conduction coefficients, etc). In this sense, temperature variations induce aging in a pile the same way moisture \cite{Bocquet1998,Bocquet2002}, constant applied stress \cite{Losert2000} and chemical reaction \cite{Gayvallet2002} do. However, the mechanism at stake lays on a pinning-depinning transition at the grain scale generated by the shear-induced successive contractions and dilations cycles. Temperature variations are thus a local and very delicate way to perturb and induce displacement in a granular assembly.

\subsection{Open questions and outlooks}

First, the role of several parameters still remains to be assessed: the mean temperature $T$ around which the variations of amplitude $\Delta T$ are imposed might play a role on the reorganization process inside the pile. Because of the presence of the container and the fact that the granular pile is a deformable medium, the mean temperature impacts on the initial stress distribution and on the space available for the grains. Its role fully remains to be investigated. Also, the parameters relevant in the case of vibrated grains are to be studied here. Namely, the shape of the grains \cite{Villaruel2000,Lumay2004,Ribiere2005b,Lumay2006}, their surface roughness \cite{Ludewig2004,Vandewalle2007}, their polydispersity \cite{Kudrolli2004}, etc. But also, the size ratio of the grains and the container \cite{Philippe2002}. Such a study might help to compare more accurately tapping and thermal cycling experiments. It could then be relevant to mix the two types of solicitation on a granular assembly to test the statistical framework proposed by Edwards \cite{Richard2005,Makse2004}. Indeed, recent experimental results lead to think that the steady state packing fraction reached by a shaken granular column (and function of $\Gamma$) is a genuine thermodynamic state, within this theoretical framework \cite{Schroter2005,Ribiere2007}. It might thus be relevant to study the fluctuations of packing fraction \cite{Nowak1998} around its equilibrium value, induced by cycles of temperature. Also, moisture is another relevant parameter that would deserve an exhaustive experimental study. Indeed, the existence of a liquid bridge between two adjacent grains is expected to increase the effective thermal conductivity \cite{Vargas2002} and thus to reinforce the role of the force chains inside the pile.

Second, the aging phenomenon observed under thermal cycling needs to be better characterized. In particular, how can we compare the aging effects observed in vibrated granular piles and described in \cite{Josserand2000} to the one described here ? Another way to put it would be to ask if we can loosen a sample by means of cycles of temperature. 

Third, more insights on the individual grain dynamics is needed in the dry case. Do the results of Slotterback and co-workers discussed in section~IV for immersed grains still hold in the dry case ? The techniques already used for vibrated dry granular assembly like X-rays tomography  \cite{Richard2003} or $\gamma$-rays absorption \cite{Philippe2002b} would certainly provide some answers to these questions.
Also, numerical simulations either simply based on successive dilation/contraction of the grains to mimic the effect of temperature, or taking into account the heat conduction through the granular media \cite{Vargas2007,Vargas2001,Vargas2002b} would spare time and unravel the key parameters in the dynamics at the grain scale.  

\begin{acknowledgements}
I am indebted to my PhD advisor, J.-C. G\'eminard, for introducing me 
to this flourishing topic and for countless enlightening discussions. 
I also thank S. Manneville and M.-A. Fardin for carefully reading 
this manuscript as well as E. Bertin and S. Ciliberto for fruitful 
discussions. 
\end{acknowledgements}


\begin{thebibliography}{100}
\bibitem{Remark2} The highest frequency accessible is roughly given by the time $\tau \equiv \sqrt{g/d}$ for a grain to fall freely under gravity over a size equal to its diameter $d$. For $d=1$~mm, 
one gets: $\tau^{-1}\simeq 100$~Hz.
\bibitem{Remark1} Note that the compaction process seems to be controlled both by the acceleration $\Gamma$ and by its duration $T=2\pi/\omega$. Thus, the relevant parameter for tapping experiments may not be $\Gamma \sim a/T^2$ , but rather $\Gamma T \sim A/T$ as claimed in \cite{Dijksmann2009}. However we discuss the results of tapping experiments in terms of $\Gamma$ as it is the paramater that has been used in the literature until now to describe the experimental observations.
\end{thebibliography}
\end{document}